\documentclass[12pt]{article}
\usepackage{graphicx}
\usepackage{amsmath}
\usepackage{amsfonts}
\usepackage{amssymb}
\begin{document}

\title{Yang-Mills theories using only extended fields (vectorial and scalar) as gauge fields}
\author{Max Chaves\\\textit{Universidad de Costa Rica, Escuela de Fisica}\\\textit{San Jose, Costa Rica}\\\textit{E-mail: mchaves@hermes.efis.ucr.ac.cr}}
\date{July 22, 2003}
\maketitle
\begin{abstract}
A few years ago H. Morales and the author introduced a type of generalized
derivative that contained both vector and scalar boson fields. Here it is
shown how to construct a full-fledged generalized Yang-Mills theory through
the introduction of \emph{extended field} multiplets. These are mixed fields
that include both a vector and a scalar part. It is shown how the standard
model of high energy physics appears naturally in a Yang-Mills theory that
uses extended field multiplets through two spontaneous symmetry breakings, one due to the VEV
of a scalar field and another to the VEV of a vector field.
\end{abstract}

About 4 years ago in a series of papers H. Morales and I showed how to write a
generalized covariant derivative that used both vector and scalar
bosons.$^{1-4}$ Some of the Lie group generators $T^{a}$ were associated with
scalar and some with vector fields. I am presenting here related but more
general results I have recently obtained, on how to build a full nonabelian
gauge theory in terms of \emph{extended fields,} which are boson fields that
have both vector and scalar components. In the old way of doing it some fields
were chosen to be vectorial and other scalar; now they are all mixed. I will
then show that the high energy standard model is the unitary gauge of one of
these theories after spontaneous symmetry breaking (SSB).

I will first give a representative example of a nonabelian gauge theory, and
then show how to \emph{generalize} it to a theory with extended fields. Take
the Lie Group to be $SU(N).$ Let $A_{\mu}^{a}$, $a=1,\ldots,N^{2}-1,$ be the
fields to be associated with the generators $T^{a}$. The fermions $\psi(x)$
are in a spinorial representation of the Lorentz group, say the fundamental
$\emph{N}$\textbf{ }of $SU(N)$. In this context we would have $A_{\mu}%
^{a}T^{a}\rightarrow UA_{\mu}^{a}T^{a}U^{-1},$ where $U$ is an element of the
fundamental representation$,$ that is, the $A_{\mu}^{a}$ transforms in the adjoint.

Treating the $A_{\mu}^{a}$ as \emph{dynamical} fields obeying a gauge
principle, we postulate the gauge transformation%
\[
A_{\mu}\rightarrow UA_{\mu}U^{-1}+U(\partial_{\mu}U^{-1})
\]
and the covariant derivative:%
\[
D_{\mu}=\partial_{\mu}+A_{\mu},\qquad A_{\mu}=igA_{\mu}^{a}(x)T^{a}%
\]
where $g$ is a coupling constant. It follows then that under the gauge
transformation the covariant derivative transforms as%
\begin{align*}
D_{\mu}f  &  =(\partial_{\mu}+A_{\mu})f\\
&  \rightarrow\partial_{\mu}f+U(\partial_{\mu}U^{-1})f+UA_{\mu}U^{-1}\\
&  =U(\partial_{\mu}+A_{\mu})(U^{-1}f),
\end{align*}
or simply%
\[
D_{\mu}\rightarrow UD_{\mu}U^{-1}.
\]

The Lagrangian of the YMT is then the obviously invariant%
\[
\mathcal{L}_{YMT}=\bar{\psi}iD\!\!\!\!/\psi+\frac{1}{2g^{2}}\widetilde
{\operatorname{Tr}}\left(  [D_{\mu},D_{\nu}][D^{\mu},D^{\nu}]\right)  ,
\]
which can be written in the more common form%
\[
\mathcal{L}_{YMT}=\bar{\psi}i(\partial\!\!\!/+A\!\!\!/)\psi+\frac{1}{2g^{2}%
}\widetilde{\operatorname{Tr}}\left(  \partial_{\lbrack\mu}A_{\nu]}+[A_{\mu
},A_{\nu}]\right)  ^{2}.
\]
The $g$ is a coupling constant. Notice that after doing all the algebra there
are no partials left acting to the right. The trace $\widetilde
{\operatorname{Tr}}$ is over the $SU(N)$ group generators, where the tilde is
used to differentiate it from the trace over Dirac matrices.

The covariant derivative has to be generalized in the spinorial
representation; this cannot be done in the vectorial representation. We need
the following theorem, whose proof can be found in Refs. 2 and 4.

\noindent\textbf{Theorem}. Let $D_{\mu}=\partial_{\mu}+B_{\mu}$, where
$B_{\mu}$\ is a vector field (either abelian or nonabelian). Then:%
\[
\left(  (\partial_{\lbrack\mu}B_{\nu]})+[B_{\mu},B_{\nu}]\right)  ^{2}%
=\frac{1}{8}\mathrm{Tr}^{2}D\!\!\!\!/^{\,2}-\frac{1}{2}%
\mathrm{\operatorname{Tr}}D\!\!\!\!/^{\,4}.
\]

Using the Theorem we can write the kinetic energy of a YMT in the spinorial
representation of the Lorentz group:%
\[
\frac{1}{2g^{2}}\widetilde{\mathrm{Tr}}\left(  \partial_{\lbrack\mu}A_{\nu
]}+[A_{\mu},A_{\nu}]\right)  ^{2}=\frac{1}{2g^{2}}\widetilde{\mathrm{Tr}%
}\left(  \frac{1}{8}\mathrm{Tr}^{2}D\!\!\!\!/^{\,2}-\frac{1}{2}\mathrm{Tr}%
D\!\!\!\!/^{\,4}\right)  .
\]

We proceed to define the \emph{extended} fields%
\begin{equation}
\Upsilon\equiv A\!\!\!/+\Phi=\gamma^{\mu}A_{\mu}+\gamma^{5}\varphi
\label{extended field}%
\end{equation}
where%
\[
A_{\mu}=igA_{\mu}^{a}T^{a},\quad\varphi=-g\varphi^{a}T^{a};\text{\quad}%
\]
the $\varphi^{a}$ are boson fields. (Alternatively we could define
$\tilde{\Upsilon}=\Upsilon^{a}T^{a}$, $\Upsilon^{a}\equiv\gamma^{\mu}igA_{\mu
}^{a}-\gamma^{5}g\gamma\varphi^{a}.)$ We require these dynamical fields to
obey the gauge transformation law%
\[
\tilde{\Upsilon}\rightarrow U\tilde{\Upsilon}U^{-1}-(\partial\!\!\!/U)U^{-1}.
\]
The generalized covariant derivative is defined to be%
\begin{equation}
D=\partial\!\!\!/+\tilde{\Upsilon}, \label{covariant derivative}%
\end{equation}
so that the covariant derivative has to transform as%
\[
D\rightarrow UDU^{-1}.
\]
Finally, we \emph{define} the gauge invariant Lagrangian of a Yang-Mills
theory with extended fields to be%
\[
\mathcal{L}=\bar{\psi}iD\psi+\frac{1}{2g^{2}}\widetilde{\mathrm{Tr}}\left(
\frac{1}{8}\mathrm{Tr}^{2}D^{\,2}-\frac{1}{2}\mathrm{Tr}D^{\,4}\right)  ,
\]
where the $\psi$ are the fermion fields.

Expanding the covariant derivative and taking the Dirac traces of this
Lagrangian one obtains:%
\begin{align}
\mathcal{L}  &  \mathcal{=}\bar{\psi}(i\partial\!\!\!/+A\!\!\!/)\psi+\frac
{1}{2g^{2}}\widetilde{\mathrm{Tr}}\left(  \left(  \partial_{\lbrack\mu}%
A_{\nu]}+[A_{\mu},A_{\nu}]\right)  ^{2}\right) \nonumber\\
&  -g\bar{\psi}i\gamma^{5}\varphi^{a}T^{a}\psi\label{expansion}\\
&  +\frac{1}{g^{2}}\widetilde{\mathrm{Tr}}\left(  \left(  \partial_{\mu
}\varphi+[A_{\mu},\varphi]\right)  ^{2}\right)  .\nonumber
\end{align}
The first line is a Yang-Mills theory; the second line is Yukawa terms; the
third is the usual kinetic energy of the scalar bosons in a Yang-Mills theory.
Notice that there are no interactions between the scalar fields higher than quadratic.

As an example we work out the theory resulting from the gauge group $SU(6)$;
it is similar to $SU(5)$ unification but with some advantages. We put the
extended gauge fields in the adjoint $\mathbf{35}$ so that the covariant
derivative is%
\[
D=\partial\!\!\!/+(\mathbf{35)}=\partial\!\!\!/+\tilde{\Upsilon}.
\]

A maximal subgroup of $SU(6)$ is $SU(5)\times U(1),$ and the obvious course is
to try to obtain the $SU(5)$ grand unified theory (GUT) through SSB. This is
the way Morales and I proceeded a few years ago. The branching rule for this
maximal subgroup is%
\[
\mathbf{35=1}_{0}\mathbf{+5}_{6}\mathbf{+\bar{5}}_{-6}\mathbf{+24}_{0},
\]
or, in a simplified symbolism,%
\[
(\mathbf{35)}=\left(
\begin{array}
[c]{cc}%
\mathbf{24} & \mathbf{5}\\
\mathbf{5}^{\dagger} & \mathbf{1}%
\end{array}
\right)  .
\]
We took the \textbf{5} to be scalar Higgs field that gives masses to the
$W^{\pm}$ and the $Z^{0}$ and the \textbf{24} to contain the gauge vectors of
the $SU(5)$ Yang-Mills.

Besides involving rather arbitrary choices of which fields to make scalar or
vector, this way of doing things has a problem that was not noticed at first.
In the usual $SU(5)$ GUT, the Higgs field that eventually generate the very
high masses for the leptoquark vector bosons are contained in another
\textbf{24.} After SSB the 12 scalar fields on the block diagonal vanish in
the unitary gauge and reappear as the third degrees of freedom of the 12 off
block diagonal vector bosons that have become massive. However, in the scheme
Morales and I used the symmetry breakings are done by Higgs fields associated
with diagonal generators of the $\mathbf{35,}$ singlets of the
$\mathbf{24\subset35}$, so there are no dynamical degrees of freedom available
to cover for the longitudinal degree of freedom the vector bosons develop
after they have become massive.

What I am proposing now is to always assume all gauge fields are extended.
That is, the gauge field associated to each generator of the covariant
derivative (\ref{covariant derivative}) must be extended in the way specified
by (\ref{extended field}). It turns out that the standard model appears if one
then assumes two SSBs. In particular, the vector part of the \textbf{5}
obtains a large mass, so that the massless Higgs structure appears in a
natural fashion.

The Lie group $SU(6)$ has 5 diagonal generators, which we, with the benefit of
hindsight, will take to be (writing the diagonal components only):%
\[%
\begin{tabular}
[c]{rrrrrrrrrr}%
$T^{1}$ & = & ( & 1 & --1 & 0 & 0 & 0 & 0 & ),\\
$T^{2}$ & = & ( & 1 & 1 & --2 & 0 & 0 & 0 & ),\\
$T^{3}$ & = & ( & 0 & 0 & 0 & 1 & --1 & 0 & ),\\
$T^{4}$ & = & ( & 2 & 2 & 2 & --3 & --3 & 0 & ),\\
$T^{5}$ & = & ( & 1 & 1 & 1 & --1 & --1 & --1 & ).
\end{tabular}
\]
This is not quite the same original choice Morales and I decided upon. We used
a different fifth generator,%
\[%
\begin{tabular}
[c]{rrrrrrrrrr}%
$T^{5}$ & = & ( & 1 & 1 & 1 & 1 & 1 & --5 & ),
\end{tabular}
\text{ (not used)}%
\]
in accord with our idea that what was involved was the maximal subalgebra
$SU(6)\supset SU(5)\times U(1).$ The choice I am giving above corresponds to
the maximal subalgebra $SU(6)\supset SU(3)\times SU(2)\times U(1)\times U(1).$
This is a full \emph{bona fide} maximal subalgebra, although it has the
peculiarity of having two $U(1)$ subgroups. It is generated by having VEVs
associated with the $T^{4}$ and $T^{5}$ generators. The crucial point here, in
order to obtain precisely the standard model out of pure extended fields, is
that one of the two fields with a VEV must be a vector field. It would not be
acceptable to have the vector field developing a nonzero VEV being one of the
vector carriers of the known forces, but, as we shall see, the vector field
with the VEV is the one associated with the generator $T^{5}$ and is not a
carrier of one of the known forces$.$ Making the assumption, usual in GUTs,
that the mass of a Higgs boson is of the order of its VEV, it would be an
extremely heavy vector boson, and thus would not be observable. This situation
raises the possibility of a small Lorentz breaking. Different such scenarios,
particularly due to high-level corrections from string theory, has been
studied in the last few years.$^{5}$

In expansion (\ref{expansion}) of the Lagrangian of the generalized Yang-Mills
theory there are two terms that allow mass generation:%
\begin{equation}
\frac{1}{g^{2}}\widetilde{\mathrm{Tr}}\left(  \left(  [A_{\mu},\varphi
]\right)  ^{2}\right)  ,\quad\frac{1}{2g^{2}}\widetilde{\mathrm{Tr}}\left(
\left(  [A_{\mu},A_{\nu}]\right)  ^{2}\right)  . \label{couplings}%
\end{equation}
We conclude from their study that a scalar boson with a VEV can generate
masses in vector fields but not in other scalar fields, and that a vector with
a VEV can generate masses in both vector fields and scalar fields.

There is a coupling between two fields when their commutator is not zero. Let
us assume that the vector part of the extended field associated with $T^{5}$
has a VEV $v$. Then the couplings of the extended fields in $\Upsilon$ with
the VEV are given by%
\[
\lbrack vT^{5},\tilde{\Upsilon}]=v\left(
\begin{array}
[c]{cccccc}%
0 & 0 & 0 & 2\Upsilon_{-} & 2\Upsilon_{-} & 2\Upsilon_{-}\\
0 & 0 & 0 & 2\Upsilon_{-} & 2\Upsilon_{-} & 2\Upsilon_{-}\\
0 & 0 & 0 & 2\Upsilon_{-} & 2\Upsilon_{-} & 2\Upsilon_{-}\\
-2\Upsilon_{+} & -2\Upsilon_{+} & -2\Upsilon_{+} & 0 & 0 & 0\\
-2\Upsilon_{+} & -2\Upsilon_{+} & -2\Upsilon_{+} & 0 & 0 & 0\\
-2\Upsilon_{+} & -2\Upsilon_{+} & -2\Upsilon_{+} & 0 & 0 & 0
\end{array}
\right)  .
\]
Thus all fields (vector or scalar) on those two corners acquire large masses.
Let us now assume that the scalar part of the extended field associated with
$T^{4}$ has a VEV $w.$ Then the coupling of this scalar field to the others in
$\Upsilon,$ as given by (\ref{couplings}), is only to their vector part. The
fields that get a mass are:%
\[
\lbrack wT^{4},\tilde{\Upsilon}]=w\left(
\begin{array}
[c]{cccccc}%
0 & 0 & 0 & 5\Upsilon_{-}^{V} & 5\Upsilon_{-}^{V} & 2\Upsilon_{-}^{V}\\
0 & 0 & 0 & 5\Upsilon_{-}^{V} & 5\Upsilon_{-}^{V} & 2\Upsilon_{-}^{V}\\
0 & 0 & 0 & 5\Upsilon_{-}^{V} & 5\Upsilon_{-}^{V} & 2\Upsilon_{-}^{V}\\
-5\Upsilon_{+}^{V} & -5\Upsilon_{+}^{V} & -5\Upsilon_{+}^{V} & 0 & 0 &
-3\Upsilon_{-}^{V}\\
-5\Upsilon_{+}^{V} & -5\Upsilon_{+}^{V} & -5\Upsilon_{+}^{V} & 0 & 0 &
-3\Upsilon_{-}^{V}\\
-2\Upsilon_{+}^{V} & -2\Upsilon_{+}^{V} & -2\Upsilon_{+}^{V} & 3\Upsilon
_{+}^{V} & 3\Upsilon_{+}^{V} & \Phi
\end{array}
\right)  ,
\]
where the superindex $V$ emphasizes that only the vector part of the extended
field is left in the matrix component.

The gauge fields that remain massless throughout both SSB are:%
\begin{equation}
\tilde{\Upsilon}_{\text{massless}}=\left(
\begin{tabular}
[c]{cccccc}\cline{1-3}%
\multicolumn{1}{|c}{$\Upsilon$} & $\Upsilon_{-}$ & $\Upsilon_{-}$ &
\multicolumn{1}{|c}{$\times$} & $\times$ & $\times$\\
\multicolumn{1}{|c}{$\Upsilon_{+}$} & $\Upsilon$ & $\Upsilon_{-}$ &
\multicolumn{1}{|c}{$\times$} & $\times$ & $\times$\\
\multicolumn{1}{|c}{$\Upsilon_{+}$} & $\Upsilon_{+}$ & $\Upsilon$ &
\multicolumn{1}{|c}{$\times$} & $\times$ & $\times$\\\cline{1-5}\cline{4-4}%
$\times$ & $\times$ & $\times$ & \multicolumn{1}{|c}{$\Upsilon$} &
$\Upsilon_{-}$ & \multicolumn{1}{|c}{$\Upsilon_{-}^{S}$}\\
$\times$ & $\times$ & $\times$ & \multicolumn{1}{|c}{$\Upsilon_{+}$} &
$\Upsilon$ & \multicolumn{1}{|c}{$\Upsilon_{-}^{S}$}\\\cline{4-5}\cline{4-6}%
$\times$ & $\times$ & $\times$ & $\Upsilon_{+}^{S}$ & $\Upsilon_{+}^{S}$ &
\multicolumn{1}{|c|}{$\Upsilon$}\\\cline{6-6}%
\end{tabular}
\right)  \label{massless}%
\end{equation}
The Higgs fields $\Upsilon_{\pm}^{S}$ are an isospin doublet of the maximal
subalgebra $SU(3)\times SU(2)\times U(1)\times U(1)\subset SU(6)$. The
$3\times3$ block is the $SU(3)$ adjoint and the $2\times2$ is the $SU(2).$ In
the description (\ref{massless}) above of the massless fields I have not
written explicitly the different diagonal fields. Let us do their bookkeeping.
The two large blocks take up three of the five extended fields along the
diagonal. The vector part of the $T^{4}$ is the usual vector boson associated
in $SU(5)$ GUTs with hypercharge (the scalar part has a nonzero VEV). The
scalar part of the $T^{5}$ decouples completely from all fields since it
commutes with them (the vector part has the other nonzero VEV), so the only
visible interacting fields are in $SU(3)\times SU(2)\times U(1)$.

The vector fields off the block diagonal have become massive and acquired an
extra degree of freedom. As it is also the situation in customary GUTs these
degrees of freedom ``eat'' the scalar bosons of the irrep that have remained
massless after the SSB. One must go to the unitary gauge in order to get rid
of these spurious fields, and these means using some of the gauge freedom to
transform to them away. In customary $SU(5)$ GUTs this is done using the gauge
freedom of $SU(5)$ that does not include $SU(3)\times SU(2)\times U(1).$ The
same thing happens here. One could think that, due to the fact that the gauge
group in our case is larger there would be additional gauge freedom, yet, this
is not true. The reason is that the Higgs fields that have appeared, I mean,
the $\Upsilon_{-}^{S},$ cannot be mixed with the other fields in the diagonal
block matrix, since the $\Upsilon_{-}^{S}$ are not fixed, but transform as a
\textbf{2} of $SU(2)$. The forecloses the use of the gauge degrees along the
sixth row and column of the transformation matrices.

Another way of understanding this same point is recalling that there are two
SSBs, and the remnant symmetry cannot mix massive particles with the massless,
for either case; again, we cannot used the gauge freedom of the sixth row and column.

In $SU(6)$ there is a $\mathbf{15=6\times6}|_{\text{antisymmetric}}$ that
contains the fermions. The term $\bar{\psi}iD\psi$ gives the correct couplings
and quantum numbers for the fermions, as been pointed out in previous papers.

The whole Standard Model can be written using only two terms and two irreps,
provided we have at our disposal the two required SSBs. No light is shed in
this model on the origin of the VEVs. The triplet-doublet problem of GUTs and
supersymmetric GUTs does not appear at all, as the Higgs \textbf{5 }are
naturally cut in two due to the way the SSBs occurred.

It is important to notice the following general point. Nowadays the standard
model is pictured as an effective low-mass quantum field theory of some
ultimate theory with a large energy scale. The masslessness of the vector
bosons of the standard model is protected by gauge invariance, and the
masslessness of the fermions is protected by chiral symmetry. But nothing
protects the scalar bosons from the large energy scale. An important function
of supersymmetry is doing precisely that, at least in the limit of small
supersymmetry breaking. In the present theory we have an alternative mechanism
to protect the masslessness of the scalar bosons. In terms of the extended
fields the theory has its own current conservation and Ward identities that
can be used to show the good behavior of scalar bosons self-energies.

\noindent\textbf{References.}

\noindent1. The idea of mixing both types of fields goes back to Y. Ne'eman,
\emph{Phys. Lett. }\textbf{B81, 190 }(1979), and D.B. Fairlie, \emph{Phys.
Lett.}\textbf{\ B82,} 97 (1979). Early work was also done by R. E.
Ecclestone,\emph{\ J. Phys}. \textbf{A13,} 1395 (1980) and \emph{Phys. Lett.}
\textbf{B116,} 21 (1982). Additional related papers can found in references 2,
3 and 4.

\noindent2. M. Chaves and H. Morales, \emph{Mod. Phys. Lett.} \textbf{A13,}
2021 (1998). M. Chaves, ``Some Mathematical Considerations about Generalized
Yang-Mills Theories'', in \emph{Photon and Poincar\'{e} Group}, ed. V. V.
Dvoeglazov, (Nova Science Publishers, New York, 1998) p. 326. M. Chaves and H.
Morales, ``The Standard Model and the Generalized Covariant Derivative'', in
\emph{Proceedings of the International Workshop ``Lorentz Group, CPT, and
Neutrinos''}, Universidad Aut\'{o}noma de Zacatecas, M\'{e}xico, June 23-26,
1999, (World Scientific) 188.

\noindent3. M. Chaves and H. Morales, \emph{Mod. Phys. Lett.} A, \textbf{15},
197 (2000).

\noindent4. M. Chaves, ``Introduction to Generalized Yang-Mills Theories'', in
``Lectures of the Summer School on Theoretical Physics, Zacatecas, M\'{e}xico,
July 31-August 5, 2000'', Hadronic J. Supplement, vol. 17, no. 1, (2002) p.
3-51, ed. V. V. Dvoeglazov y A. E. Mu\~{n}oz.

\noindent5. V. Kostleck\'{y} and S. Samuel, Phys. Rev. \textbf{D39 }(1989)
683; Phys. Rev. \textbf{D40 }(1989) 1886; Phys. Rev. Lett. \textbf{63} (1989)
224; Phys. Rev. Lett. \textbf{66} (1991) 1811.
\end{document}